\documentclass[preprint,aps,showpacs,showkeys,amsmath,amssymb]{revtex4}

\usepackage{graphicx}
\usepackage{dcolumn}
\usepackage{bm}
\usepackage{amsmath}
\usepackage{amsfonts}
\usepackage{amssymb}
\usepackage{amsthm}
\usepackage{newlfont}
\usepackage{graphicx}
\usepackage{amscd}

\begin{document}


\title{Motzkin numbers out of Random Domino Automaton}

\author{Mariusz Bia{\l}ecki}
\email{bialecki@igf.edu.pl}
 
\affiliation{%
Institute of Geophysics, Polish Academy of Sciences \\
ul.~Ks. Janusza 64, 01-452 Warszawa, Poland}%

\date{\today}

\begin{abstract}
Motzkin numbers are derived from a special case of Random Domino Automaton  
- recently proposed toy model of earthquakes \cite{BiaCzAA}. 
An exact solution of the set of equations describing stationary state
of Random Domino Automaton in "inverse-power" case is presented. 
A link with Motzkin numbers allows to present explicit form of asymptotic behaviour
of the automaton.  
\end{abstract}

\pacs{
02.30.Lt (Sequences, series, and summability),  
02.10.Ox (Combinatorics; graph theory),
02.50.Ey (Stochastic processes), 
45.70.Ht (Avalanches), 
}
\keywords{stochastic cellular automata, Motzkin numbers, avalanches, solvable models, integer sequences, toy models of earthquakes}

\maketitle


It is known, how to obtain Catalan numbers out of the bond directed percolation on a square lattice \cite{Inui-Catalan}.
Here we derive Motzkin numbers \cite{FlaSedComb, Aigner, SloaneEnc, SloaneOE} from the recently proposed
Random Domino Automaton  \cite{BiaCzAA, CzBiaTL, CzBiaEF}, which may be regarded as a toy model of earthquakes.

Random Domino Automaton comes from very simplified view of earthquakes. 
The space - one dimensional lattice - corresponds to boundaries of two tectonic plates moving with relative constant velocity. Due to irregularities of surfaces, relative motion can be locked at some places producing stress accumulation.
Beyond some threshold of the stress, a relaxation took place. The size of relaxation depends on the nearby accumulated stress.  
Energy in the automaton is represented by balls added to the randomly chosen cell (each one is equally possible) 
with a constant rate - one ball in one time step. 
If the chosen cell is empty, it becomes occupied with probability $\nu$ or the ball is scattered with probability $(1-\nu)$ leaving the state of the automaton unchanged. 
If the chosen place is already occupied, there are also two possibilities: the ball is scattered with probability $(1-\mu)$ or with probability $\mu $ the incoming ball triggers a relaxation - balls from the chosen cell and all adjacent occupied cells are removed. 
An example of relaxation of size {\bf five} is presented in the diagram below.
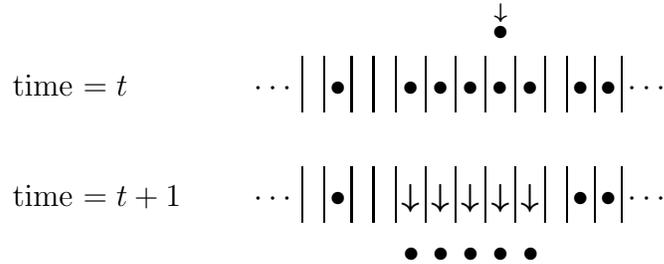
\begin{center} 
		\begin{tabular}{l c|c|c|c|c|c|c|c|c|c|c|c|c|c}
		\multicolumn{9}{c}{ \ } &  \multicolumn{1}{c}{$  \stackrel{\boldsymbol{\downarrow}}{\bullet} $}&  \multicolumn{5}{c}{\ }		\\
		 time $= t $   &  $\quad \quad \cdots$ & $ \ $ & $ \bullet $ & $ \ $ & $ \ $ & $ \bullet $ & $ \bullet $ & $\bullet$ &
			$ {\bullet} $ & $  \bullet $ & $ \ $ & $ \bullet $ & $ \bullet $ &$ \cdots \quad \quad $	   \\	
\multicolumn{15}{c}{ \ }		\\
		 time $= t + 1 $  &  $ \quad \quad \cdots$ & $ \ $ & $\bullet$ & $ \  $ & $ \ $ & $\boldsymbol\downarrow$ & $\boldsymbol\downarrow $ &
$\boldsymbol\downarrow $ & $\boldsymbol\downarrow$ &  $\boldsymbol\downarrow$ & $ \ $ & $\bullet$ & $ \bullet $ & $\cdots \quad \quad$    \\
\multicolumn{6}{c}{ \ } &  \multicolumn{1}{c}{$ {\bullet} $} &  \multicolumn{1}{c}{$ {\bullet} $}& \multicolumn{1}{c}{$ {\bullet} $} & 
\multicolumn{1}{c}{$ {\bullet} $} & \multicolumn{1}{c}{$ {\bullet} $} & \multicolumn{4}{c}{\ }		\\
			\multicolumn{15}{c}{ \ }		\\
		\end{tabular} 
\end{center}

The stationary state of the system may be described by the distribution of clusters.
The number of clusters of the length $i$, for $i=1,2,\ldots$, is denoted by $n_i$; the number of empty clusters
of length $1$ is denoted by $n_1^0$. Then the number of all clusters $n$ and and the density $\rho$ are  
\begin{equation}
n= \sum_{i\geq1} n_i, \quad  \quad \rho = \frac{1}{N}\sum_{i\geq1} i n_i.
\label{eq:nandrho}
\end{equation}
The following set of equations is derived from the stationarity conditions (see \cite{BiaCzAA})
\begin{eqnarray} 
n_1&=&\frac{1}{\frac{\mu_1}{\nu}+2}\left((1-\rho)N  - 2 n + n_1^0 \right), \label{eq:n1}\\
n_2 &=& \frac{2}{2\frac{\mu_2}{\nu}+2} \left( 1- \frac{n_1^0}{n} \right) n_1 , \label{eq:n2} \\
 n_i &=& \frac{1}{\frac{\mu_i}{\nu} i+2} \times   \nonumber \\
 && \times \left( 2 n_{i-1}  \left( 1- \frac{n_1^0}{n} \right) 
+ n_1^0 \sum_{k=1}^{i-2} \frac{n_k n_{i-1-k}}{n^2} \right) \label{eq:ni}  
\end{eqnarray}
for $i\geq 3$,
where 
$$n_1^0  = \frac{ 2 n}{\left(3 + \frac{2\sum_{i\geq1} \mu_i i n_i}{\nu n}\right)}.$$
From the above set the balance equation for  the total number of clusters $n$  and for the density $\rho$ can be derived
\begin{equation}
(1-\rho){N} - 2n  =  \sum_{i\geq1} \frac{\mu_i}{\nu} n_i i,
\label{eq:ballance_n}
\end{equation}
\begin{equation} \label{eqn:equil_ni}
\nu(1-\rho) = \frac{1}{N} (\sum_{i\geq1} \mu_i n_i i^2).
\end{equation}

In the case which refers to equal probability of triggering an avalanche for each cluster, 
the parameters are fixed as follows: $\nu=\text{const}$ and
$\mu=\frac{\delta}{i}$, where ${\delta=\text{const}}$.
Equations \eqref{eqn:equil_ni} and \eqref{eq:ballance_n} give
$ \rho=(\theta+1)^{-1}$, and 
$ n=N\theta[(\theta+1)(\theta+2)]^{-1}$,
where $\theta=\frac{\delta}{\nu} \in (0,\infty)$.
Together with simple form of $n_1^0  = {2 n}/{\left(3 + 2\theta \right)}$ 
it allows to  reduce the set of equations \eqref{eq:n1}-\eqref{eq:ni} to the following recurrence:
\begin{eqnarray} 
n_1&=&\frac{N \theta}{(\theta+1)(\theta+2)^2}\left( \theta + \frac{2}{(2\theta+3)} \right), \label{eq:Bn1}\\
n_2 &=& \frac{2}{\theta+2} \left(\frac{2\theta+1}{2\theta+3} \right) n_1 , \label{eq:Bn2} \\
n_{i+1} &=& \frac{2}{(\theta+2)}\left(\frac{2\theta+1}{2\theta+3} \right) n_{i}  + \nonumber \\
 && + \frac{2}{N\theta} \left( \frac{\theta+1}{2\theta+3} \right) \sum_{k=1}^{i-1} {n_k n_{i-k}}  \label{eq:Bni}  
\end{eqnarray}
for $i\geq2$.

We define new variables  $c_i$ for $i = 0,1,\ldots,$ by 
\begin{equation} \label{eq:Cidef}
c_i = \frac{\beta}{\alpha^{i+1}}n_{i+1},
\end{equation}
where
\begin{eqnarray}
\alpha &=& \frac{2}{(\theta+2)}\left(\frac{2\theta+1}{2\theta+3} \right), \\
\beta &=& \frac{(\theta+1)(\theta+2)}{N\theta(2\theta+1)}.
\end{eqnarray}
Then, the equation \eqref{eq:Bni} can be rewritten in the form 
\begin{equation} \label{eqn:rec}
c_{m+2}=c_{m+1} + \sum_{k=0}^{m} {c_k c_{m-k}},  
\end{equation}
which is valid for $m \geq 0$ ($m=i-2$). Initial data $c_0$ and $c_1$ are easily obtained 
from equations \eqref{eq:Bn1}-\eqref{eq:Bn2}, when it is transformed according the rule of equation \eqref{eq:Cidef},
namely
\begin{equation}
c_0 = c_1  = \frac{1+\frac{3}{2}\theta + \theta^2 }{1+ 4 \theta + 4 \theta^2}.
\label{eq:c}
\end{equation}

The above equation \eqref{eqn:rec} has the form of Motzkin numbers recurrence \cite{FlaSedComb, Aigner, SloaneEnc, SloaneOE}. 
It is similar to the ubiquitous Catalan numbers recurrence 
(however the order of \eqref{eqn:rec} is not $1$ but is $2$) 
and can be solved using generating functions technique \cite{FlaSedComb}. 
For the limit case, when $\theta=0$, we start with $c_0=c_1=1$, and recurrence \eqref{eqn:rec}
produces Motzkin numbers: $1, 1, 2, 4, 9, 21, 51, 127, 323, 835, 2188, \ldots$, etc.
Thus, it is explicitly shown, how to obtain them from the Random Domino Automaton.

Below we present solution of the set \eqref{eq:Bn1}-\eqref{eq:Bni}.
The generating function 
\begin{equation}\label{eq:genfg}
C(z)=\sum_{m\geq0} c_m z^m 
\end{equation}
for the recurrence \eqref{eqn:rec} is equal to
\begin{equation} 
\frac{(1-z) - \sqrt{1-2z +(1-4c_0)z^2 + 4(c_0-c_1)z^3 }}{2z^2}  \label{eq:genfs}.
\end{equation}
For $c_0=c_1$  it reduces to
\begin{equation}
G(z)= \frac{(1-z) - \sqrt{1-2z(1+kz)}}{2z^2}
\label{eq:gfz}
\end{equation}
where $k= (2c_0 -\frac{1}{2}).$
From expansion of the formula \eqref{eq:genfs} one can read explicitly the form of coefficients $c_m$ being 
solutions of the recurrence \eqref{eqn:rec}. 
Using 
\begin{equation}
(1+z)^a = \sum_{n\geq0} {a \choose n} z^n,
\label{eq:expans}
\end{equation}
where 
\begin{equation}
{a \choose n} = \frac{a(a-1)(a-2)\ldots(a-n+1)}{n!},
\label{eq:choose}
\end{equation}
for $m\geq0$ we have
\begin{equation}
[ z^m ] gf(z) = [z^{m+2}] \left( -\frac{1}{2} \sqrt{1-2z(1+kz)} \right).
\label{eq:zm}
\end{equation}
Then it follows
\begin{eqnarray*}
 \sqrt{1-2z(1+kz)} = \sum{n\geq0} {\frac{1}{2} \choose n} (-2)^n z^n (1+kz)^n =   \\
 = 1-z-kz^2  + \sum_{n\geq2} \frac{-1}{n 2^{n-2}} {2n-3 \choose n-1} z^n (1+kz)^n. 
\label{eq:expsq}
\end{eqnarray*}
For any $m\geq0$ it gives
\begin{equation}
c_m=[z^{m+2}] \left( \frac{kz^2}{2} + \frac{1}{2} \sum_{n\geq2} \frac{1}{n 2^{n-2}} {2n-3 \choose n-1} z^n 
 \sum_{j=0}^n {n \choose j} k^j z^j \right).
\label{eq:cma}
\end{equation}
Changing indices $n+j=m+2$ for $m\geq1$ we have 
\begin{equation}
c_m=\frac{1}{2} \sum_{j=0}^{[\frac{m+2}{2}]} \frac{(2c-\frac{1}{2})^j}{(m-j+2)2^{m-j}} 
{ 2(m-j)+1 \choose m-j+1 } { m-j+2  \choose j} 
\label{eq:cmfin}
\end{equation}
Thus formula \eqref{eq:Cidef} gives explicit solution of equations \eqref{eq:n1}-\eqref{eq:ni}
for the distribution $n_i$s for any value of $\theta$.

Using known assymptotics for Motzkin numbers (after Benoit Cloitre, see \cite{SloaneOE})
\begin{equation}
c(i) \sim \sqrt{3/4/\pi} \frac{3^{i+1}}{i^{\frac{3}{2}}}
\label{eq:motzkin}
\end{equation}
we can present explicit formula for asymptotic behaviour for the $n_i$ distribution. 
In the limit case $\theta \longrightarrow 0$, while $N \theta = const$,
\begin{equation}
n_i =  \frac{\alpha^{i}}{\beta} c_{i-1} \longrightarrow (N \theta) (\frac{1}{3})^i c_{i-1},
\label{eq:nic}
\end{equation}
and using \eqref{eq:motzkin} one obtains
\begin{equation}
n_{i+1} \sim \frac{1}{i^{\frac{3}{2}}}.
\label{eq:invpowni}
\end{equation}
Thus we found one more example of inverse-power distribution with the power equal to $\frac{3}{2}$.

Author would like to thank prof. Zbigniew Czechowski for fruitful discussions.

\end{document}